# Modelling African swine fever virus spread in pigs using time-respective network data: scientific support for decision-makers

Mathieu Andraud[1,*], Pachka Hammami[1], Brandon Hastings Hayes[2], Jason Ardila Galvis[3], Timothée Vergne[2], Gustavo Machado[3], Nicolas Rose[1]

[1]ANSES, EPISABE Unit, Ploufragan-Plouzané-Niort Laboratory, Ploufragan, France

[2]UMR ENVT-INRAE IHAP, National Veterinary School of Toulouse, Toulouse, France

[3] Department of Population Health and Pathobiology, College of Veterinary Medicine, Raleigh, NC, USA

*Corresponding author: mathieu.andraud@anses.fr

## Summary

African Swine Fever (ASF) represents the main threat to swine production, with heavy economic consequences for both farmers and the food industry. The spread of the virus that causes ASF through Europe raises the issues of identifying transmission routes and assessing their relative contributions in order to provide insights to stakeholders for adapted surveillance and control measures. A simulation model was developed to assess ASF spread over the commercial swine network in France. The model was designed from raw movement data and actual farm characteristics. A metapopulation approach was used, with transmission processes at the herd level potentially leading, through a reaction-diffusion process, to external spread to epidemiologically connected herds. Three transmission routes were considered: local transmission (e.g. fomites, material exchange), movement of animals from infected to susceptible sites, and transit of trucks without physical animal exchange. Surveillance was based on prevalence and mortality detection thresholds, which triggered control measures based on movement ban for detected herds and epidemiologically related herds. The time from infection to detection varied between 8 and 21 days, depending on the detection criteria, but was also dependent on the types of herds in which the infection was introduced. Movement restrictions effectively reduced the transmission between herds, but local transmission was nevertheless observed in higher proportions highlighting the need of global awareness of all actors of the swine industry to mitigate the risk of local spread.

Raw movement data were directly used to build a dynamic network on a realistic time-scale. This approach allows for a rapid update of input data without any pre-treatment, which could be important in terms of reactivity, should an introduction occur.

## Keywords

Transmission model, Infectious Diseases, Epidemiology, Surveillance, Control

## Introduction

Infectious diseases represent a major concern for livestock production due to economic consequences and related societal impacts as well as veterinary and public health issues. In the last decade, African swine fever (ASF) became the main concern for the European and Asian swine industry and more recently the North American one because of its reintroduction to the Caribbean (Amat et al., 2021; Gonzales et al., 2021; Wang et al., 2021). Following its introduction into Georgia in 2007, ASF has gradually spread in Europe and is now considered endemic in some countries (Chenais et al., 2019; Pejsak et al., 2018). To date, only two countries that had been infected with the ASF virus have regained their free status: Belgium and the Czech Republic (Danzetta et al., 2020; Emond et al., 2021). Recently, an introduction of the virus into the wild boar population at the German-Polish border was detected (Sauter-Louis et al.). At the moment of the study, Novembre 2021, this epidemics was still ongoing with geographical expansion in Germany despite the implementation of control measures. Although limited to wildlife for about 10 months after introduction, the first case in domestic pigs was registered in July 2021 (Federal Ministry of Food and Agriculture, 2021). Understanding the spreading patterns of infectious agents at local and global scales is necessary to design relevant control measures (Beaunée et al., 2015). The identification of pathogen transmission routes and the assessment of their relative role in the transmission process are essential (Andronico et al., 2019; Ezanno et al., 2020). Although contacts between individuals and contributions from the environment are critical for pathogen transmission at farm level, additional contact structures — such as animal movements and indirect contacts through transportation or geographical/social proximity between actors — were highlighted as pivotal in the study of transmission of animal diseases at a large geographical scale (Büttner et al., 2016; Galvis et al., 2021).

Social network analysis has the potential to shed light on contact structures, which, joined to epidemiological data, provided insights into the role of specific actors in the transmission process for livestock infections (Hardstaff et al., 2015; Salines et al., 2017; Schulz et al., 2017). Metapopulation models of epidemics were also developed to study the role of animal movements on infection spread based on either real movements or summaries derived from movement datasets (Beaunée et al., 2015; Nickbakhsh et al., 2013). Focusing on transmission models in pigs, most of those were dedicated to regulated diseases such as foot-and-mouth disease, classical swine fever and more recently ASF (Andraud & Rose, 2020). The development of generic modelling frameworks (NAADSM, ISP, DTU-DADS, Be-Fast) was highly promoted to assess the impact of control strategies using unified frameworks (Bates et al., 2003b; Boklund et al., 2009; Harvey et al., 2007; Martínez-López et al., 2013). These models account for direct between-farm (animal movement) and indirect transmission processes between epidemiological units such as farms, with anthropogenic, vectorial or environmental origins such as material sharing, feeding, vehicles used in the transportation of animals or feed, or airborne dissemination (Bates et al., 2003a; Boklund et al., 2008; Halasa, Botner, et al., 2016). However, the integration of movement data requires a pre-treatment of network data to summarize the contact rates between epidemiological units according to their intrinsic characteristics (*e.g.* herd types and sizes) and relational characteristics (*e.g.* distances and frequency of contacts according to herd types) (Büttner et al., 2016; Halasa, Bøtner, et al., 2016; Hu et al., 2017).

The conjuncture of ASF in Europe demonstrates the permanent risk of introduction of the virus into disease-free areas and the need for predictive tools to assess both the consequences such an introduction would have and the responsiveness of the surveillance protocols. Based on social network structure, a previous study demonstrated the interest of a mathematical modelling approach for targeted surveillance towards at-risk herds among the swine commercial network in France, where the network was summarized based on movement frequency and distances between production sites (Andraud et al., 2019). The present study comes to complete this modelling framework through the integration of realized movement data, represented on a realistic time scale. An original metapopulation model was developed, fed with raw movement data, and parameterized according to ASF epidemiological specifications, to assess the extent to which ASF would spread among domestic pig holdings in France, as well as the effectiveness of surveillance and control strategies.

## Material and Methods

### Data

Data were extracted from the national swine movement database BDPORC (Rautureau et al., 2012; Salines et al., 2017). Briefly, notification of all shipments of live animals is compulsory in France since 2010, allowing for tracing of contacts between swine production units. The origin, the destination, the date, the number of animals that were exchanged and the truck identification number are collected at every shipment. Farm specific data are also available, allowing for geolocation and informing on the herd types and capacities (total number of animals). The data used hereafter cover the period from 2017 to 2019, over which two types of networks were designed, namely the animal introduction model (AIM) and the transit model (TM) (Figure 1) [22]. While the AIM approach considers effective contacts through the movements of live animals between production sites, the transit model accounts for complex rounds consisting of successive loadings and unloadings in different sites. The order of visited farms during the rounds was informed by the database, allowing for reproducing the exact contact tracing. TM links all herds that were visited by the same truck during a round. In the AIM, movements to slaughterhouses and rendering plants were neglected as they were considered as epidemiological endpoints (Rautureau et al., 2012). However, if several herds were visited to collect slaughter-aged animals during the same round and using the same truck, they were fully considered in the TM due to the possibility of viral transmission from one herd to the other.

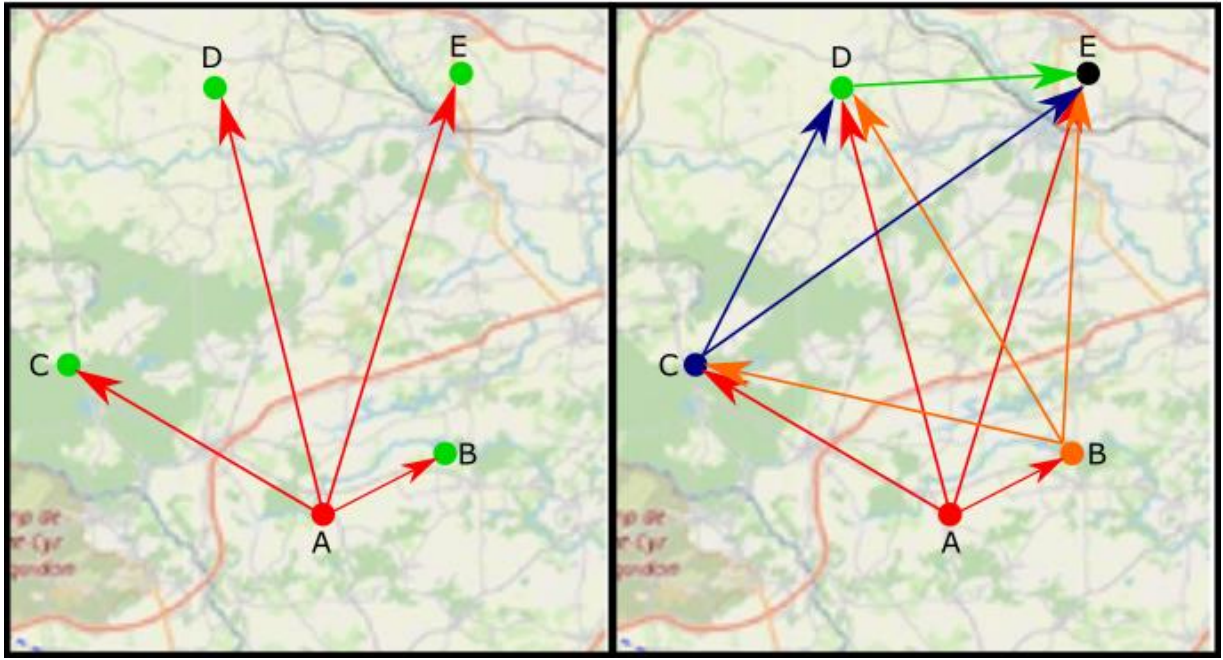

**Figure 1.** Illustration of the Animal introduction Model (AIM; left) and the Transit Model (TM; right) Network designs. The movements are in sequential order with a first loading of animals in farm A and subsequent unloadings of subgroups of animals in farms B, C, D and E. The AIM considers effective animal movements with 4 herds and 4 directed links during the same round and with the same truck (from A to other sites, red lines). In contrast, the TM considers the truck passage without animal exchange between sites, linking site B to C, D and E (orange lines), Site C to D and E (blue lines), and site D to E (green line).

## Epidemiological model

A multiscale model was designed to account for the within-herd transmission process, which governs in turn the transmission to commercial partners assuming three different transmission routes: local transmission affecting neighbouring herds (farm-to-farm proximity), transmission through the AIM, and indirect transmission related to TM.

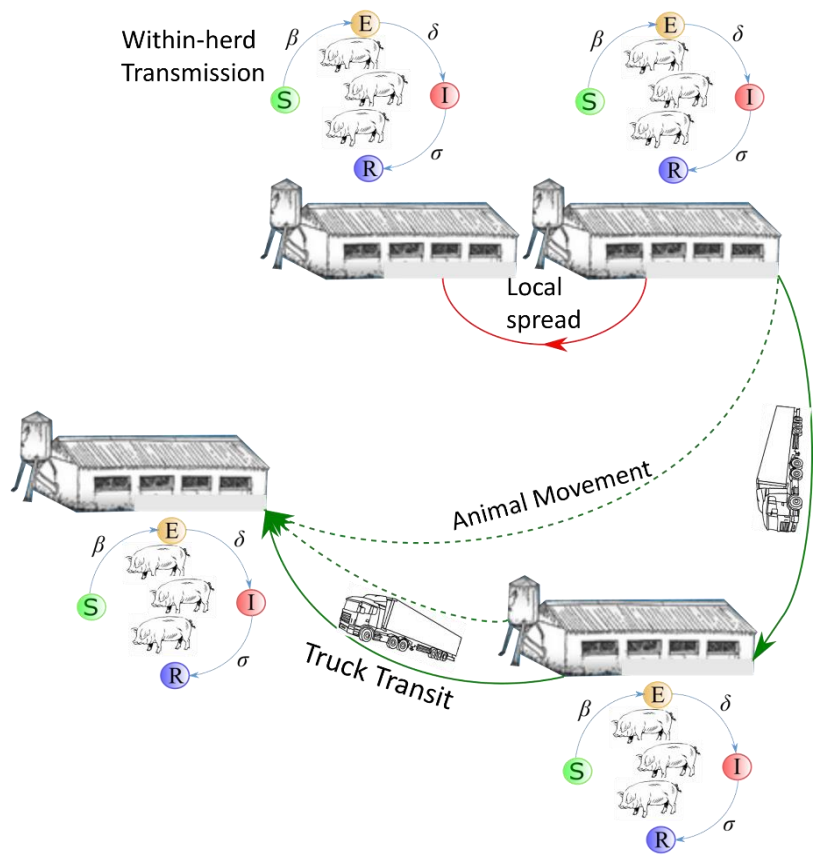

**Figure 2.** Multi-scale epidemiological units and transmission processes. Within-farm a stochastic Susceptible-Exposed-Infected-Removed (SEIR) transmission model in a homogenous population; animal movement (dashed arrows), truck transit (solid green arrows) and local transmission (solid red arrows) routes on the commercial pig trade network.

## Within-herd transmission dynamics

A stochastic Susceptible-Exposed-Infected-Removed (SEIR) model was used to represent the dynamics of infection in every infected herd. The initial number of susceptible animals was determined from the total pig capacity provided in the database, under the assumption of homogenous mixing. Running the simulations over a relatively short period of time (c.f. *Simulations* subsection described later), no birth-death process was considered except for infection-related mortality. Therefore, for agents with a high lethality rate such as ASF virus, the within-herd population decreases as the infection progresses, thereby affecting the transmission process, due to the frequency dependent formulation of the force of infection. At each time step $t$ (day), the number of exposed animals in herd $j$, $E_j^{(t)}$, is updated using a binomial process acting on the number of susceptibles $S_j^{(t)}$ with probability $\pi_j^{(t)} = 1 - exp\left(-\beta \frac{I_j^{(t)}}{N_j^{(t)}}\right)$, where $\beta$ denotes the transmission rate and $I_j^{(t)}$ and $N_j^{(t)}$, the numbers of infectious and total number of animals in herd $j$ at time $t$. Transitions from exposed to infectious, and from infectious to removed stages are modelled in the same way, with an exponentially distributed infectious period of mean $\delta$ and $\sigma$, respectively (transition rates: $1/\delta$, $1/\sigma$). When lethal infections are considered, mortality was assumed to occur at the end of the infectious period, with a proportion $\mu$ of deaths among the number of newly removed animals. A herd was defined as infected when at least one pig was infected, initiating the within-herd transmission process.

## Between-herd transmission dynamics

Three transmission routes were considered for the spread of the infectious agent between production sites.

*Local transmission*

The transmission in the neighborhood of infected herds may have different origins, *e.g.* airborne transmission, indirect contact through shared material, passage of vehicles, human factors, as well as fomites or vector transmission (*e.g.* hematophagous flies), which can be non-negligible routes of infection for ASF (Bonnet et al., 2020; Mauroy et al., 2021; Vergne et al., 2021). A distance-dependent transmission rate $\beta_{Loc}(d_{i,j})$ was considered using a Gaussian kernel centered on every infected farms within a radius of $D = 2\sigma$, $\sigma$ being the standard error of the Gaussian kernel (expressed in meters). The force of infection was represented as proportional to the number of infected individuals in the infected herds ($I_i$). Therefore, the animals housed in a susceptible farm $j$, located at a distance $d_{i,j}$ from an infected herd $i$, were exposed to a force of infection $\lambda_{Loc}^{(j)}(t) = \beta_{Loc}(d_{i,j}) * I_i^{(t)}$, leading to a probability of infection at each time step equal to $\pi_{Loc}^{(j)}(t) = 1 - exp\left(-\lambda_{Loc}^{(j)}(t)\right)$. As described for within-herd transmission, the introduction process is governed by a binomial process at each time step, leading or not to the infection of some animals in the surrounding herds. Once a herd is infected via this route, meaning that at least one animal is infected and initiates the within-herd transmission process, local transmission is then assumed negligible for this herd.

*Network driven transmission*

In this study, raw data were directly integrated to parameterize the movements respecting their origin and destinations, as well as their actual timing, daily contact networks.

Animal introduction model (AIM)

A reaction-diffusion approach was considered to represent the spread of the agent between farms. For each recorded shipment sourced from an infected farm, the transferred individuals are randomly drawn from the total population of the farm (herd size) according to the number of animals actually moved recorded in the database, and regardless of their infection status. They are then integrated into the target population (replacing former susceptible animals to keep the population size constant) and replaced by susceptible individuals in the source population. A susceptible farm was considered infected when at least one infectious individual was introduced, initiating the within-herd transmission process.

Transit model

In addition to the reaction-diffusion approach considered for animal movements, when a transport vehicle passes through an infected farm, a probability $\pi_{TM}$ of transmission of the agent to farms downstream in the round (dashed lines in Figure 1), in the absence of unloading of animals, was considered. This probability was assumed constant, independent from herd sizes or the numbers of farms visited during the round, using a Monte-Carlo algorithm to characterize the occurrence of infection events. When a farm was infected through this route, only one individual was shifted from susceptible to an exposed epidemiological state, its transition to the infectious stage initiating the within-farm transmission process.

## Monitoring and control measures

### Detection

The detection module, due to passive surveillance, was based on two indices reflecting the evolution of the infection at the herd level: a prevalence threshold, defined by the proportion of infected individuals in the farm, and a mortality threshold, defined by the cumulative number of dead (or removed) individuals in the farm (Table 1). In the absence of demographics, only pathogen-related mortality was considered. When a positive herd was detected, control measures were triggered and surveillance was enhanced by lowering the detection thresholds to notification of every dead animal and a prevalence of infected individuals of 1%.

### Surveillance

In case of regulated diseases, the model evaluates the number of herds to be monitored based on the French emergency plan (DGAL/SDSPA/2019-41; DGAL/SDSPA/2019-41). When a case is detected, surveillance and protection zones (SZ and PZ) are set up for 40 days within a 10- and 3-km radius, respectively. All herds in these two zones are monitored during that period. Mortality notification is compulsory and systematically induces clinical inspection leading to further laboratory analyses (virology and serology). Downward and upward tracings are also performed to identify the contacts due to animal movements that occurred up to 30 days prior to detection.

### Control measures

Implemented control measures were based on movement restrictions. Detected herds were depopulated and considered as not infectious anymore by any transmission route. A complete containment was implemented for herds identified in surveillance and protection zones as well as those traced as potential contacts of the infected herd, similarly to what would be implemented in the event of an outbreak.

## Simulations

### General Considerations

The parameter set used for simulation purposes is provided in table 2. For a matter of illustration, these parameters were selected in ranges matching to ASF epidemiological knowledge and were mainly derived from the literature (Guinat et al., 2016; Halasa, Bøtner, et al., 2016). Based on the pyramidal structure of the swine industry in France (Rautureau et al., 2012; Salines et al., 2017), the herds were distributed in three categories corresponding to nucleus and multiplier herds (further called "Breeders"), farrowing-only and post-weaning production farms ("Farrowers"), and farms having a finishing sector ("Finishers"). For each of these farm categories, 1,000 simulations were run, each with a randomly selected seeding herd within the given category. Each simulation could last up to 40 days, but was stopped as soon as there were no more infected herds in the system. To minimize the risk of stochastic extinction in the first days of simulations, the number of infectious animals in the index herd was set to five. Thus, simulations were initialized at a random date (in the 3 years of data) through the introduction of 5 infected individuals in a random index herds, selected among those having at least one animal movement in the 40 days of simulation.

**Table 1.** Parameters of the epidemiological model.

| Parameter | Definition (unit) | Value (Reference) |
|---|---|---|
| $\beta$ | Between-individuals Transmission rate (day$^{-1}$) | 0.6 (Guinat et al., 2016) |
| $\beta_{loc}$ | Maximum Local Transmission rate (day$^{-1}$) | 0.0001 (Halasa, Botner, et al., 2016) |
| $D$ | Radius for local transmission kernel (m) | 500 |
| $\beta_{TM}$ | Transit transmission probability | 0.05 (Expert opinion) |
| $\delta^{-1}$ | Latent period (day) | 4 (Halasa, Botner, et al., 2016) |
| $\gamma$ | Infectious period (day) | 7 (Halasa, Botner, et al., 2016) |

*Scenarios*

The baseline scenario consisted of no-control spread. Although unrealistic for a regulated disease such as ASF, this scenario demonstrated the extent to which an infectious agent could spread over the network when considering realized shipment data.

*Detection and movement restriction*

The initial detection thresholds were set to 5% of infected individuals at the herd level, along with the death of 5 individuals (Table 2). Based on that scenario, the prevalence threshold was then decreased to 1% in scenarios 3, followed by a decrease of the mortality threshold down to 1 animal in scenario 4. After the first detection, the threshold were set to scenario 4 values to represent the onset of enhanced surveillance protocol.

**Table 2.** Parameters for detection of the first herd initiating control measures and enhanced surveillance. After the first detected case, the parameters are set to those of scenario 4.

| | | Detection parameters | |
|---|---|---|---|
| Scenario | detection | Prevalence of infected animals Threshold | Mortality Threshold (number of dead pigs) |
| 1 | FALSE | - | 0 |
| 2 | TRUE | 0.05 | 5 |
| 3 | TRUE | 0.01 | 5 |
| 4 | TRUE | 0.01 | 1 |

*Outcomes*

For each scenario and seeded introduction at each herd category, the number of simulations where secondary infections occurred was recorded. For those simulations, the number of secondary infected herds —along with the transmission route and the distance to the source of infection — were registered. Where applicable, the delay between infection and detection, otherwise stated as the 'at risk' period, for each infected and detected herd was also analyzed.

The model was developed in R software, using the igraph package to represent the time-respective networks, rgeos, rgdal and sf to deal with spatial features and shiny for user-friendly implementation interface. The model is available on GitHub, along with a data sample for testing (https://github.com/mandraud/EpiNet/tree/main).

# Results

*Network descriptions*

Animal introduction model

The network accounts for 12,397 active herds located in France and involved in 268,672 shipments over the three years of data (2017-2019). Considering shipments with loadings and unloadings of animals between production sites (figure 3A and 3B), the average number of animals per shipment was 150 pigs (2.5 – 97.5 percentiles: 1 - 672) and were dependent on the category of the source herd. Indeed, shipments from nucleus farms mostly involved small groups of animals, with a median size of 34 animals (range [1 – 280]), while those from farrowing and post-weaning herds concerned full batches with a median group size about 208 piglets ([8 – 830], Figure 3B). However, the highest out degree was reached by nucleus herds with a median number of 195 outgoing contacts ([2 – 908], Figure 3A) during the three years. Only 32% of herds with finishing barns were involved in outgoing shipment with unloadings of post-weaned animals in other finishing herds, and they had very limited outgoing contact rate (median outdegree 12 [1 – 124]).

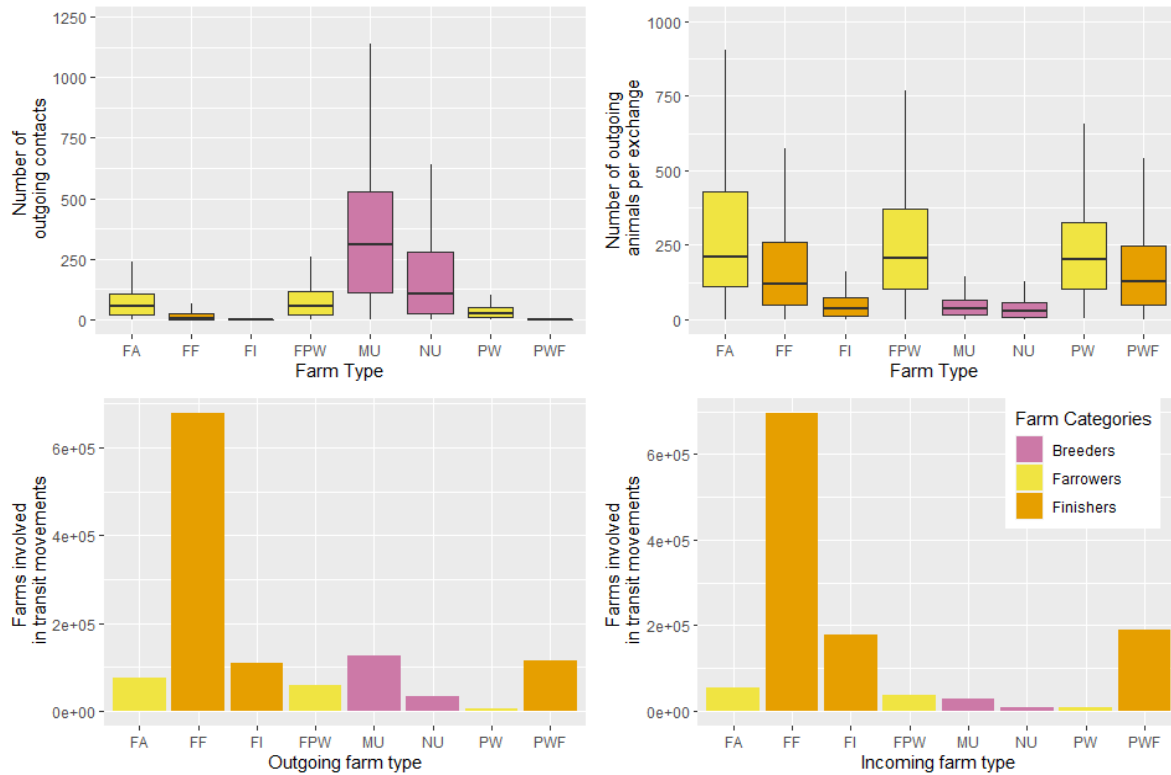

**Figure 3.** Network characteristics for animal movements (AIM, panels A and B) and transit (TM, panels C and D) from 2017 to 2019 for the different herd types in France (FA: Farrowing, PW: Post-Weaning, FPW: Farrowing-Post-Weaning, PWF: Post-Weaning-Finishing, FI: Finishing, FF: Farrow-to-Finish, MU: Multipliers, NU: Nucleus). A. Number of outgoing contacts (loadings); B. Number of animals exchanged per movement; C. Number of transit movement from each farm type; D. Number of transit movement to each farm type.

Accounting for truck transit in rounds dramatically densifies the contact structure with up to 916,271 links in the network involving 12,795 production sites in 238,159 rounds. This increase is obviously due to extensive rounds visiting high numbers of farms. The role of herds with finishing sectors was found essential, covering 75% of sites involved in the TM network due to sequential loadings from different farms to slaughterhouses (Figure 3C and 3D).

Based on network characteristics, three groups of herd types with similar characteristics were designed for further analysis: Farrowers corresponding to herds with farrowing-only and/or post-weaning sectors (represented in yellow on boxplots in figure 3), Finishers, to herd having a finishing sectors (orange on boxplots in figure 3), and Breeders, to multiplier and nucleus herds (purple on boxplots in figure 3). The color scheme remains consistent throughout the paper.

*Simulations*

*Scenario 1:* "No-control" Baseline situation

The theoretical results in the absence of control measures are described in Figure 4. On average, about 70% of the simulations resulted in secondary cases. Transmission was quasi-systematic when the introduction was set in breeders and farrowers (79 and 82% of simulations; Figure 4A) compared to finishers (49%). Introductions in Breeder herds led to major outbreaks with a median of 12 secondary cases, but potentially rising up to 52 cases (Figure 4B). In contrast, introductions in other production herds (farrowers or finishers) seeded the infectious agent with a lower intensity, affecting a median number of 3 herds (95% interval: [3 – 12]); the distance ranges were also two-times lower reaching 200 km (Figure 4C). Direct contact represented the major route of transmission of the virus, between 68% and 85% of transmission events. This was followed by transmission through indirect contact in the neighbourhood which, despite a fairly small radius of interaction (500 m), represented 14 to 29% of events depending on the category of farm of introduction. Finally, transit movements (TM) were implicated as being responsible for 2% of the cases recorded during the simulations (Figure 4D). Transmission through the movement of animals or vehicle transit occurred mainly on a relatively small geographical scale, with median distances ranging between 10 and 30 km. Some movements nevertheless generated more distant infections, which could reach 400 km when the virus was introduced in breeders.

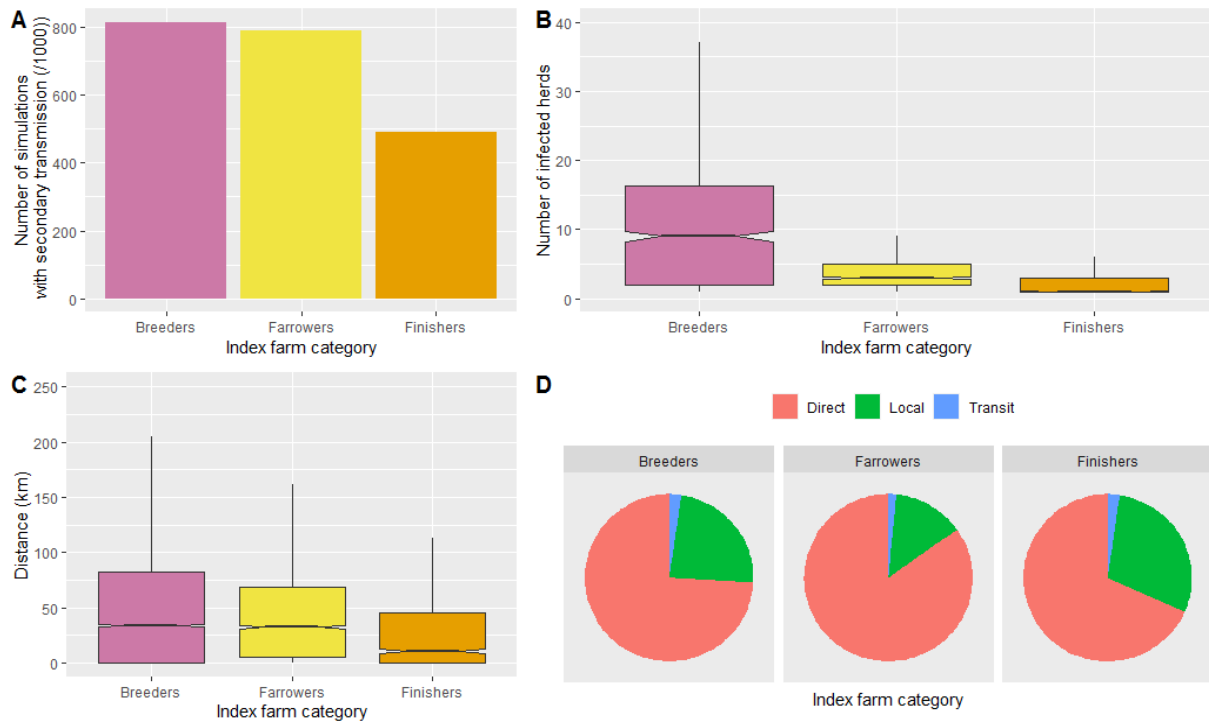

**Figure 4.** Simulation results in the absence of control measures according to the category of index farm by the end of 40 days of disease propagation: A. number of simulations leading to secondary transmissions (among 1000 simulations); B. number of infected farms; C. distances between source farms and secondary cases; D. relative contribution of different transmission routes (direct for pig movements, local and transit related transmissions)

Control measures

*Scenario 2. Detection criteria: 5% Prevalence threshold and 5 dead animals*

The detection process, based on a prevalence of 5% of infected animals and a disease-related mortality of 5 animals per farm, induced a median time to detection of 14 to 21 days post-infection depending on the category of infected unit (Figure 5A-E). All detection events were effective within 30 days following infection, supporting the duration of the contact-tracing period. Farrowers were detected earlier than breeders and finishers, due to lower herd sizes leading to earlier fulfillment of detection criteria. Finishers were the most affected farm category with up to 4743 detection events over the whole set of simulations (3000 iterations, Figure 5E).

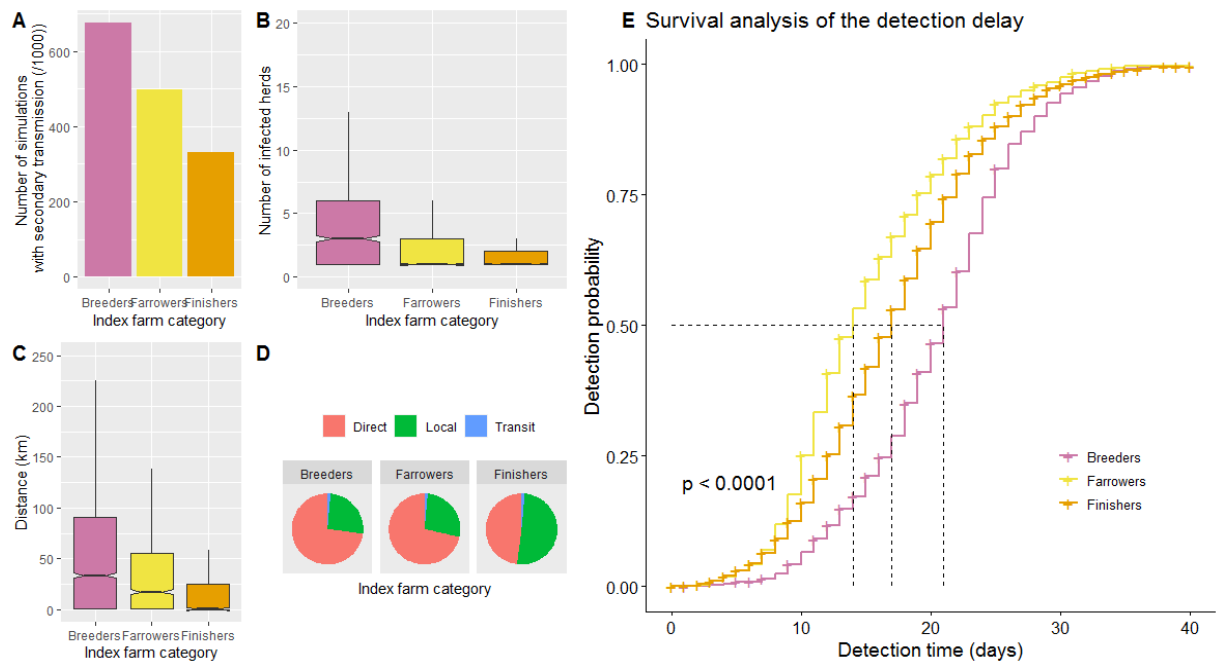

**Figure 5.** Analysis of detection probability according to time since infection depending on the farm category (scenario with prevalence and mortality detection thresholds of 5% and 5 animals per farm, respectively). A. Number of simulations leading to secondary transmissions (among 1000 simulations); B. Number of infected farms; C. Distances between source farms and secondary cases; D. Relative contribution of different transmission routes (direct for pig movements, local and transit related transmissions); E. Survival analysis of the delay of detection.

During the 'at-risk' period, defined as the delay between infection and detection; the occurrence of secondary infections was reduced by 17% when breeders were selected as the initial seeders (675 simulations) with transmission chains limited to 5 herds ([2 – 20], Figures 5A and 5B). Detection and subsequent control measures reduced the number of secondary infections by 28% and 32% when seeding infection in finishers and farrowers, respectively, but did not affect the length of transmission chains when transmission occurred, with a median of 2 to 3 infected farms per run. The overall impact of the control measures induced a reduction of the total number of infections due to animal transfer, making the relative contribution of local transmission route more important, representing 26% to 27% of transmission events when the virus was introduced in breeders and farrowers, and becoming the major transmission route when introduction occurred in finishers (50.6%, Figure 5D). The median distance between new cases and their source of infection was evaluated to 66.5 km for direct transmission when seeding in breeders (43 and 32 km for farrowers and finishers, respectively, Figure 5C).

*Scenario 3. Detection criteria: 1% Prevalence threshold and 5 dead animals*

Decreasing the prevalence of infected pigs' threshold for detection to 1% led to a faster response to the transmission process (Figure 6A-E). The median time to detection was evaluated from 11 to 14 days for all farm types. The earliest detections were still observed when introduced in farrowers, starting on day 6 post-infection, but differences between herd categories were reduced (Figure 6E). Although the detection time was significantly decreased when primary infection was seeded in farrowers and breeders, the impact on the number of simulations with effective transmission when starting from finishers was not significant. In contrast, breeders were detected much earlier when lowering the prevalence detection threshold (21 days in scenario 2, 13 days in scenario 3), confirming the importance of the out-degrees in regards to the transmission process.

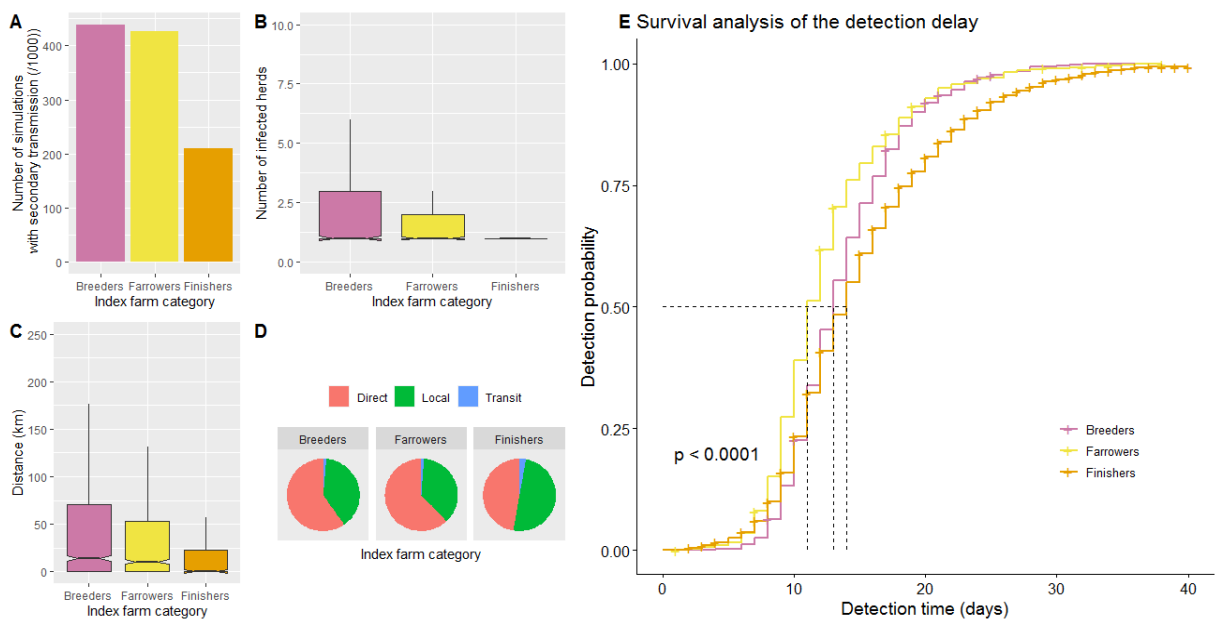

**Figure 6.** Simulation results under control scenario 3 (for prevalence and mortality detection thresholds of 1% and 5 animals) according to the type of index farm. A. number of simulations leading to secondary transmissions (among 1000 simulations); B. number of infected farms; C. distances between source farms and secondary cases; D. relative contribution of different transmission routes (direct for pig movements, local and transit related transmissions); E. survival analysis of the delay of detection (3000 simulations).

The proportion of simulations with effective transmission to secondary cases fell to 21% when the agent was introduced in finishers, and was limited to 43% in the alternative introduction schemes (Figure 6A). A median of one secondary outbreak was observed wherever the introduction occurred, but breeders could spread the infection up to 13 herds due to an extended "at-risk" period and important outgoing contact chains (Figure 6B). A movement ban after the first detection drastically reduced the number of infected herds, which were mostly due to local transmission thereafter (Figure 6D).

*Scenario 4. Detection criteria: 1% Prevalence threshold and 1 dead animal*

Decreasing the mortality threshold to 1 animal improved the detection in simulations when the infection was seeded in farrowers (median time to detection estimated to 7 days, Figure 5A-E). It remained unchanged in breeders and finishers. This was directly reflected by the proportion of simulations leading to secondary infections which fell to 33% when infection was introduced in farrowers (Figure 7A). None of the evaluated control measures were found to inflect the distance pattern for transmission processes (Figures 5-7C), despite the higher proportion of local transmission events.

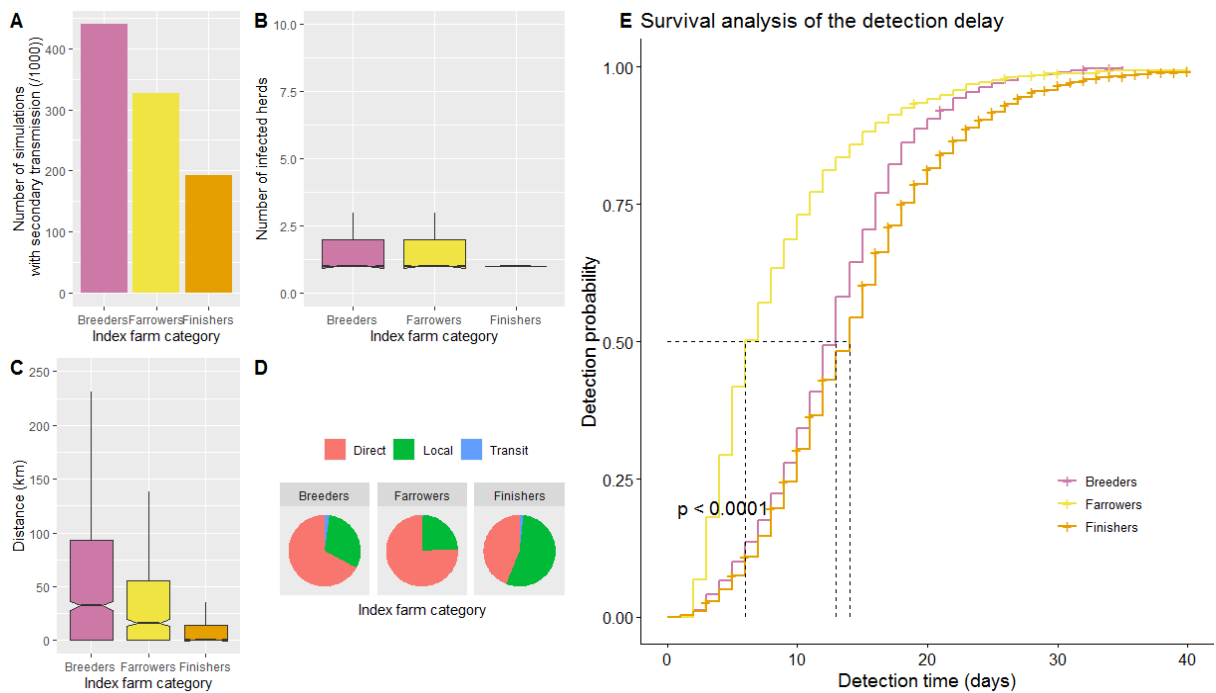

**Figure 7.** Simulation results under control scenario 3 (for prevalence and mortality detection thresholds of 1% and 1 animal) according to the type of index farm. A. number of simulations leading to secondary transmissions (among 1000 simulations); B. number of infected farms; C. distances between source farms and secondary cases; D. relative contribution of different transmission routes (direct for pig movements, local and transit related transmissions); E. survival analysis of the delay of detection (3000 simulations).

## Discussion

African swine fever represents a major threat for the swine industry and related economic sectors. Preparedness to face such a threat is therefore a need in countries that are still free of the disease. The development of predictive tools, based on risk assessment or mechanistic models, offers scientific support to decision makers. Several mechanistic models were developed to tackle potential introduction and dissemination of ASF given the increased risk of global spread — as highlighted with its expansion throughout Europe and Asia and now into the Caribbean (Gonzales et al., 2021). A recent review identified four main objectives for these studies (Hayes et al., 2021). The present study comes at the crossroad of three of those through the evaluation of the contribution of the transmission routes, the assessment of the epidemiological consequences of a hypothetical outbreak and the assessment of effectiveness of control measures. Based on farms' characteristics and movement of animals, the results highlighted the relationship between the transmission potential of the herds and their position in the swine pyramidal production system. A quantitative evaluation of the consequences of pathogen introduction in swine trade network was conducted. Although live pig movements were identified as the main transmission route, the results evidenced the need of compliance with external biosecurity, to limit the transmission in the neighborhood of infected farms as well as disinfection of vehicles.

Here we developed a metapopulation model to represent the on-farm spread of the pathogen combined with a transmission process between production sites based on movement data. Such an approach was already described in several studies and revealed the huge potential of multiscale modelling (Andraud & Rose, 2020; Bastard et al., 2020; Beaunée et al., 2015). The work of Beaunée *et al.* on *Mycobacterium avium subsp. paratuberculosis* (Map) spread in cattle clearly illustrated the advantage of integrating information at different scales to have an exhaustive view of the epidemiological context (Beaunée et al., 2015). Including data at the farm level along with records of animal movements allowed analyzing the conditions for introduction and persistence of Map on farms and at a regional scale. Dedicated to ASF in the domestic pig compartment, Halasa et al. proposed a multiscale approach, combining a within-herd transmission model with a reconstruction of the contact network based on the frequency of contacts registered between herd types and distances between production sites (Halasa, Bøtner, et al., 2016). More recently, Machado et al. designed a model representing ASF spread in Brazil, taking 3-by-3km cells of a geographical raster as epidemiological units and representing movements between these units (Gustavo Machado et al., 2021).

Three transmission routes were considered for between-sites transmission: the animal movement pathway remains dominant but as soon as movement restriction occurs through the control measures, the relative contribution of local transmission increases. This is under the assumption that there is no modification of local biosecurity measures in case of an outbreak, which may be somewhat unrealistic. Hence, an evolution of local biosecurity, through awareness of all local actors, appears as a key element for disease control. Although playing a minor role in the transmission process, truck transit should not be neglected and complex rounds serving several production sites have to be considered in the tracking surveillance system.

The delay from infection to detection, or the 'at-risk' period, was analyzed depending on the sensitivity of detection in infected farms. Using different prevalence and mortality thresholds, the time to detection varied between 8 and 21 days, in line with results obtained from Halasa's model (Andraud et al., 2019 - Submitted; Halasa, Bøtner, et al., 2016). During this period infected herds may transmit the virus to epidemiologically related herds: the shorter this at-risk period, the lower the number of infected herds. The detection delay was also highly dependent on the herd types (breeding, farrowing or finishing herds). Due to their position in the pyramidal structure of the swine production system, the probability of first detecting finishing herds is rather low and the modification of detection-triggering parameters did not affect their detection delay (15 to 17 days). This suggests that such a threshold should be adapted according to both the herd characteristics (*e.g.* the type of herds defining their contact structure) and their neighbouring context, based on the local density of farms as well as their connectivity to other sites. Although the awareness of the local actors (farmers, veterinarians) is essential to limit disease spread, high levels of vigilance —in regards to social and societal aspects — would be difficult to maintain for an extremely long period of time, requiring adjustments according to the level of risk.

Highly connected herds were identified as potential key spreaders in this study, especially nucleus and multiplier herds, which have the longest outgoing contact chains. Conversely, finishing herds — with a high number of ingoing connections — appeared more vulnerable but were not at a high risk of spread because of their low number of outgoing connections. It is worth noting all herds were assumed to have similar internal and external biosecurity levels. This shortcut may create an overestimation of the transmission process, especially originating from nucleus and multiplier herds which are likely to observe strict biosecurity protocols limiting the risk of introduction of pathogens (Silva et al., 2019). The incorporation of data-based classification of herds' biosecurity levels would improve the model, providing for better guidance for surveillance targets. In the absence of such biosecurity qualifications, the choice was made to set a baseline level for all herds representing the most severe epidemiological situation.

In contrast with models based on summary statistics of movement data (Andraud et al., 2019; Halasa, Bøtner, et al., 2016), raw movement data were directly used to build a dynamic network on a realistic time-scale. This approach allows for a rapid update of input data without any pre-treatment, which could be important in terms of reactivity, should an introduction occur. However, the results remain based on historical data, which may raise the question of prediction quality in the event of an actual introduction. One solution to this issue may rely on the generation of simulated networks, with similar characteristics. In that view, Exponential Random Graph Models (ERGMs) offer a promising perspective (Kukielka et al., 2017; Phoo-ngurn et al., 2019; Relun et al., 2016). This statistical tool allows for characterizing the main factors governing the network structure based on its structural features and the node characteristics, from which simulated networks can be generated. This way, the model would get free from the past, improving outcome predictiveness.

Our model focuses on transmission dynamics of an infectious agent on swine production networks after the introduction of ASF in the domestic pig compartment. The introduction in France is nevertheless very likely to occur in wild boars, as was the case in several countries (Chenais et al., 2019; Podgórski et al., 2020; Sauter-Louis et al.). To analyze the risk of introduction into the domestic pig reservoir, the interface between wild-boar and domestic pig compartments needs to be represented (Hayes et al., 2021). This could induce an external force of infection, especially on free-range herds which were already highlighted as critical point for surveillance (Andraud et al., 2019 - Submitted).

In this study, we focused exclusively on control measures based on movement restriction and immediate depopulation of infected herds after detection. In the meantime, movement ban was implemented in surveillance and protection zone. In that view, the modelling framework could be used to evaluate alternative surveillance and control measures (*e.g.* sizes and duration of zones) to optimize their epidemiological and economic impact (G. Machado et al., 2021). The identification of 'at-risk' herds from farms and network characteristics could help designing preventive strategies, reducing so the infectious pressure in the neighborhood of infected farms, especially on high density areas (Jurado et al., 2018).

In conclusion, a transmission model of infectious agents on the swine commercial network was developed to analyze the effectiveness of surveillance and control measures. Fed with raw movement data, the modelling framework is flexible with the transmission process represented on a time-respective schedule. Applied to ASF, this model identified animal movements as the pivotal transmission route. However, local transmission and mechanical transmission through truck transit should not be neglected, and requires a global awareness of all actors from the production system.

## Acknowledgements

We would like to thank the members of BDPORC for making holdings and animal movement data available.

## Conflict of interest statement

The authors declare no conflict of interest.

## Data availability statement

Code and example datasets are made available on GitHub: https://github.com/mandraud/EpiNet/tree/main.